\documentclass[11pt]{article}
\pdfoutput=1
\usepackage{url}
\usepackage{authblk}
\usepackage{graphicx}
\graphicspath{ {./img} }
\usepackage{booktabs}

\providecommand{\keywords}[1]
{
    \small
    \textbf{\textit{Keywords---}} #1
}

\title{Qibitz: Mining PubMed for Repurposable Drugs}
\author[$\dag$]{David Massart, PhD}
\author[$\ddag$]{Marc Zeicher, MD, PhD}
\affil[$\dag$]{D.E.Solution, Brussels, Belgium}
\affil[$\ddag$]{Targeted Therapies Research \& Consulting Centre (TTRCC), Brussels, Belgium}
\date{\today}

\begin{document}

    \maketitle

    \begin{abstract}
        PubMed’s current search interface makes it tedious to systematically search for medical and research literature
        on drugs that could potentially be used to treat a given pathology, including patients with genetically altered
        tumors.

        This is because physicians must search separately for each drug-pathology combination (or drug-gene combination).
        To streamline this process, this paper proposes adding a faceted search interface to PubMed. Faceted search is a
        common feature on e-commerce websites that allows users to filter search results by selecting different fields.
        By incorporating this technology, not only can physicians save time and improve the accuracy of their literature
        searches, but also presenting search results in this way makes patterns emerge, which can suggest new
        treatment options for a given pathology (including patients with genetically altered tumors).
    \end{abstract}

    \newpage
    \tableofcontents
    \newpage
    \section*{Introduction}
    \addcontentsline{toc}{section}{Introduction}
    PubMed is a free online database that provides access to tens of millions of biomedical and life sciences
    publications~\cite{PubMed2023}, including clinical trials, observational studies, and systematic reviews~\cite{PubMedGuide2023}.
    It is maintained by the National Center for Biotechnology Information (NCBI) at the U.S. National Library of Medicine
    (NLM), and enables healthcare professionals, researchers, and students to easily search for relevant biomedical
    literature.

    PubMed indexes a variety of fields~\cite{PubMedFields2023}, including the title of the article, the abstract,
    author information, and Medical Subject Headings (MeSH terms).
    MeSH terms are standardized descriptors from the MeSH thesaurus~\cite{MeSH2023}, and all PubMed articles are indexed
    with them to ensure consistent retrieval of related articles across searches.
    This makes PubMed a primary resource for anyone searching for relevant biomedical literature.

    Physicians can use PubMed to learn about the potential interactions between a particular drug and a given medical
    (pathological or physiological) condition and between a particular drug and given pathological or physiological
    pathways networks, (including specific genes).
    To do this, they can look up the corresponding MeSH descriptors for the drug and the medical condition, and then
    search for articles in PubMed that contain these two descriptors.
    For example, if physicians want to learn about the potential interactions between the paclitaxel drug and gene
    FBXW7, they look up the MeSH descriptors D017239 and C513273, respectively.

    Suppose the physicians are interested in identifying drugs that have the potential to interact with a particular
    gene.
    They first need to compile a list of all existing drugs and their corresponding MeSH descriptors.
    They can then perform a search for each gene/drug pair of descriptors.
    If there are multiple genes of interest, the process needs to be repeated for each gene.
    This greatly increases the workload and makes the task more challenging.

    This paper proposes to address the problem of searching for drug-pathology and drug-gene interactions in PubMed by
    equipping the catalog with a faceted search interface~\cite{Tunkelang2009}.
    Faceted search is a common feature on e-commerce websites that allows users to quickly explore and refine search
    results by selecting different fields, or ’facets’ to filter the results.

    This approach makes it easier for users to learn about different drug-pathology or drug-gene interactions and
    quickly access the relevant papers describing these interactions.

    This paper is divided into sections that describes the development of a faceted search interface for PubMed.
    This interface is designed to facilitate the identification of drugs that may improve the treatment of various
    pathologies, including tumors with specific genetic alterations.
    Section~\ref{sec:example} presents a real-life use-case that motivates the practical interest of such a tool.
    The following couple of sections provide technical details:
    Section~\ref{sec:indexes} outlines the process of deriving specialized gene and drug fields from PubMed's MeSH and
    free-text fields, while Section~\ref{sec:workflow} describes the data acquisition workflow put in place to acquire
    the PubMed catalog data, enrich it, and keep it up-to-date in an automated way.
    Finally, in Section~\ref{sec:examples}, we illustrate the effectiveness of our tool and method through three
    practical real-life examples.

%
%
    \section{A Motivational Example}
    \label{sec:example}

    The gene/drug example mentioned in the introduction is not merely theoretical but has practical significance.
    FoundationOne~\cite{FoundationOne2023} is a next-generation sequencing (NGS) genetic test that examines a patient’s
    tumor to identify genetic mutations that may be driving the cancer’s growth.
    The test analyzes a large panel of genes commonly associated with cancer~\cite{Frampton2013}, using advanced genomic
    technologies.
    Physicians often request this test to assist in making treatment decisions, as it provides valuable insights into
    which targeted therapies or immunotherapies would be most effective in treating the patient’s specific cancer type.
    By personalizing treatment plans for their patients, physicians can potentially improve outcomes.
    Many pharmaceutical drugs, although they were not developed and marketed for that purpose, can interact with genes
    through the regulation of gene expression, either by activating or inhibiting gene expression~\cite{Heguy1995}.
    Many drugs’ effects on gene expression have been studied, and the results of these studies are now published in
    research papers referenced by PubMed.

    As a result, PubMed can be used to search for medications that, although they have not been specifically developed
    to treat cancers, are known to inhibit the expression of mutated genes found in a patient’s tumor.
    These drugs, if known to have no or minimal side effects, can be administered in addition to the recommended
    treatment to increase its efficiency.

    This process of finding alternative therapeutic uses for existing drugs is referred to as drug
    repurposing~\cite{Ashburn2004}, and it can be a valuable approach to improve the treatment of cancerous tumors
    presenting mutated genes.
    Search tools such as faceted search interfaces, which accelerate the exploration of scientific literature, can
    expedite the identification of drugs that have demonstrated efficacy against the expression of specific genes and
    that are therefore good candidates for drug repurposing.

    Cancers can exhibit complex genetic landscapes, with up to 500 genes becoming dysregulated.
    This dysregulation often occurs gradually, spanning 20-30 years or more, before the cancer's symptoms become apparent.
    As a result, therapeutic strategies that target only a single gene product or cell signaling pathway are unlikely to be
    effective in preventing or eradicating cancer.

    Chemotherapy and specific targeted drugs have been developed to disrupt these gene products or pathways, thereby
    inducing cell death and impeding progression of malignant changes in cells.
    However, problems such as ineffective targeting and drug resistance have plagued these agents, necessitating changes
    in the approach to systemic cancer therapy.
    The current paradigm of cancer chemotherapy is either combinations of several drugs or a drug that modulates
    multiple targets.
    The combination chemotherapy approach uses drugs with different mechanisms of action to increase cancer killing.
    Various drugs that modulate multiple targets, have been approved by the U.S. Food and Drug Administration (FDA) for
    treatment of various cancer types.
    However, these drugs are costly, have a long list of undesirable side effects, and most of them are still not
    effective enough to have a significant effect on the course of the disease.

    When one or more growth pathways are blocked, a cell, especially a cancer cell, often switches to unaffected
    parallel or compensating pathways.
    Multiple interconnected signaling pathways that promote cell growth and inhibit cell death have been identified,
    and these pathways often function differently in cancer cells compared to normal cells.
    To address this complexity, new multi-ingredient pharmaceutical compositions are needed for cancer treatment.
    By simultaneously targeting multiple pathways, these targeted therapies can be more effective.

    In practice, this goal is difficult to achieve for several reasons: new drugs are extremely expensive, their full
    pattern of side-effects and interference is incompletely known, and it is difficult to get different drug companies
    to cooperate in combination clinical trials.
    Moreover, regulatory authorities (FDA and EMEA) require that a new drug proves its efficacy in monotherapy clinical
    trials and hence synergy of two drugs, which on their own are ineffective, is missed.

    Many currently marketed drugs (for other indications than cancer) and nutraceuticals have been reported in in-vitro
    screening of drug libraries, preclinical cancer animal models epidemiological data (incidence, survival), clinical
    trials (prevention, add-on therapy) as potentially interfering with these pathways.

    Therefore, it is useful to research literature looking for already marketed non-cancer treatment related drugs
    for which exists data or evidence that they might block or inhibit one of these identified cytotoxicity
    circumvention pathways.

    As explained in the introduction, PubMed’s current search interface makes it tedious to systematically search for
    medical and research literature on drugs that could potentially be used to treat a given pathology, including
    patients with genetically altered tumors, because it forces physicians to search separately for each drug-pathology
    or drug-gene combination.
%
%
    \section{Specializing Fields for Improved Filtering Capabilities}
    \label{sec:indexes}

    Faceted search is a technique that enhances traditional search methods with a faceted navigation system, allowing
    users to refine their search results by applying multiple filters based on faceted classification of catalog
    entries~\cite{Tunkelang2009}.

    A good facet requires a field to possess the following properties:
    \begin{itemize}
        \item Relevance: The field must contain relevant information that helps users precisely narrow down their
        search results.
        \item Distinctiveness: The field must have distinct and meaningful values that can categorize search results
        into exclusive groups without overlap.
        \item Completeness: The field must contain complete and accurate information for most catalog entries,
        ensuring optimal retrieval (i.e., good recall) of filtered search results.
        \item Consistency: The field's values must be consistent across all catalog entries to accurately and reliably
        filter search results.
    \end{itemize}

    As of this writing, PubMed contains over 37 million entries, each with a unique identifier (PMID) and additional
    information such as article title, author(s), journal name, publication date, abstract, keywords, and more.
    Over time, 66 fields have been used to describe and classify the various document types referenced in
    PubMed~\cite{PubMedFields2023}.

    However, none of these 66 fields meet the criteria for creating effective facets for drugs or genes.
    This is because gene and drug information is scattered across multiple fields, making these fields incomplete and
    unreliable for faceted search.
    Moreover, no single field is dedicated exclusively to genes or drugs; instead, this information is mixed with other
    concepts, contaminating potential facets with irrelevant data.

    In short: None of these fields meet the necessary criteria for faceted search and they cannot be used as-is as a
    facet to organize search results by drugs or genes.


    \subsection{Deriving a Drug Field}
    \label{ssec:drugs}
    PubMed stores drug information across multiple fields.
    The drug name can be found in any of the following free-text fields: Title (TI), Abstract (AB), or Other Term (OT).
    The Other Term field contains free keywords assigned by indexers, as well as author-assigned keywords since 2013.
    Additionally, two controlled vocabularies can be used to identify a drug:
    \begin{enumerate}
        \item MeSH descriptors, which appear in either the MeSH Terms (MH) field or the more specialized Substance Name
        (NM) field.
        \item Registry numbers, which are unique identifiers consisting of numerical or alphanumeric codes assigned by the
        National Library of Medicine (NLM) based on chemical structure, and are stored in the dedicated Registry Number
        (RN) field.
    \end{enumerate}

    In this context, to conduct an exhaustive search for articles about, for example, aspirin, you need to use the
    following search strategy:
    \begin{itemize}
        \item Search for ``aspirin'' and its synonym ``acetylsalicylic acid'' in the free-text fields: Title (TI),
        Abstract (AB), and Other Term (OT).
        \item Search for the registry number ``R16CO5Y76E'' in the Registry Number (RN) field.
        \item Search for the MeSH descriptor ``D001241'' in the MeSH Terms (MH) field and the more specialized
        Substance Name (NM) field.
    \end{itemize}

    Having drug information scattered across multiple PubMed fields is not the only challenge.
    Each of these fields also contains non-drug related information, making it difficult to create a reliable drug facet.
    The main issue is that using any of these fields would result in poor recall and relevance.
    For instance, building a drug facet based on the MeSH Terms (MH) field would:
    \begin{itemize}
        \item Miss relevant articles that haven't been indexed with MeSH descriptors (bad recall).
        For example, an article that contains ``aspirin'' in the title but hasn't been indexed with descriptor ``D001241''.
        \item Display irrelevant values (poor relevance).
        For example all the non-drug related MeSH descriptors present in the result set, rather than limiting the list
        to drug-related descriptors (poor relevance).
    \end{itemize}

    To build a reliable drug facet that meets the criteria presented above, you need to create a controlled vocabulary
    of drugs and use it as the basis for a new drug field.
    For practical reasons (our organization is based in Belgium and we want to focus on drugs prescribed in Belgium),
    we selected the Belgian database of authorized medicines for human use (BelMed~\cite{FAMHP2023}) as the foundation
    for the drug vocabulary.

    To populate this drug field, we followed these steps:
    \begin{itemize}
        \item For each entry in the BelMed database, we chose a unique token, typically the name of the molecule
        (e.g., acetylsalicylic acid).
        \item We then created three mappings to link each token (e.g., acetylsalicylic acid) to its corresponding:
        \begin{itemize}
            \item Drug name, synonyms, and common brand names (e.g., aspirin);
            \item MeSH descriptor (e.g., D001241);
            \item Registry number (e.g., R16CO5Y76E).
        \end{itemize}
        \item Once these mappings are ready, they can be applied to the entire PubMed catalog\footnote{The PubMed database can
        be obtained in XML format from the anonymous FTP server of the National Library of
        Medicine~\url{ftp://ftp.ncbi.nlm.nih.gov}.} to enrich its records with the new drug field.
    \end{itemize}
    This enables the derivation of a reliable drug facet from the values found in PubMed's existing drug fields.

    \subsection{Deriving a Gene Field}
    \label{ssec:genes}

    In PubMed, gene information is scattered across multiple fields, similar to drug information.
    This fragmentation makes it challenging to build a reliable gene facet.
    Among these fields, the Gene Symbol (GS) field, which is specific to gene information, could have been a good
    candidate for building a gene facet if it hadn't been limited to indexing records from 1991 to 1995.
    The other fields that contain gene information combine it with non-gene information, making them unsuitable for
    building a reliable gene facet.
    Thus, as with the drug facet, it is necessary to create a controlled vocabulary of genes and using it as the basis
    for a new gene field that can serve as a robust and accurate gene facet.

    The Human Genome Organisation's (HUGO) Gene Nomenclature Committee (HGNC) is responsible for maintaining the
    standardized HUGO Gene Nomenclature.
    This committee assigns a unique gene name and symbol, comprising a combination of letters and digits, to each known
    human gene~\cite{HGNC}.
    The HGNC database, available at~\url{https://www.genenames.org}, serves as a controlled vocabulary repository.
    For practical purposes, we have chosen a subset of the HUGO gene nomenclature, specifically covering genes included
    in the FoundationOne test, as the basis for our gene vocabulary.
    Each of these genes is represented by its corresponding HGNC symbol, which serves as a unique token (e.g., FBXW7).

    The PubMed records were enriched with a new multivalued field named `genes' that was populated by extracting and
    mapping data from other PubMed fields.

    \subsubsection{Populating the Gene Field by Mapping MeSH Descriptors}

    Many genes (not all) have been assigned a MeSH descriptor, which are referenced in the MeSH Terms field (MH). To
    leverage this existing information, a mapping was created to convert these MeSH descriptors (for example, D000075924)
    into their corresponding HGNC symbols (for example, ATRX).

    \subsubsection{Populating the Gene Field by Mapping Registry Numbers}

    Some genes are assigned registry numbers, listed in the Registry Number field (RN).
    To utilize this information, we created a mapping to convert these registry numbers (e.g., 138415-26-6) into their
    corresponding HGNC symbols (e.g., PRDM1).
    However, a single registry number can sometimes correspond to multiple genes, with, in some cases, up to ten or more
    genes associated with it.
    This ambiguity severely impacts the precision of searches.
    Therefore, we opted not to rely on registry numbers as a source of gene information.
    Fortunately, nearly all articles indexed with registry numbers are also annotated with MeSH descriptors.
    As a result, this decision had a limited impact on the recall of searches.

    \subsubsection{Extracting HGNC Symbols from Unstructured Text Fields}

    The HUGO Gene Nomenclature Committee's standardized symbols are widely adopted in medical literature and appear
    unchanged in PubMed's free-text fields, such as Title (TI), Abstract (AB), and Other Term (OT).

    Many of these symbols are composed of distinctive combinations of letters (both uppercase and lowercase) and digits,
    such as C11orf30 and DNMT3A. These symbols are unique enough to be accurately and reliably identified within a text
    without requiring any additional processing.

    However, short symbols consisting only of capitalized letters (four letters or fewer) are ambiguous because
    they often correspond to acronyms commonly used in other medical domains.
    When a sequence of capitalized letters matching one of these ambiguous symbols is found, the number of false
    positives (i.e., the number of matches that do not correspond to a gene symbol) can be reduced by checking the
    surrounding text to see if it discusses genetics (i.e., contains variations of terms like ``genes'' and ``genetics'').
    This is enough to disambiguate some symbols.
    For example, this method is usually sufficient to distinguish between VHL, the symbol of the gene coding for the von
    Hippel-Lindau tumor suppressor and VHL the acronym of the Virtual Health Library.

    However, there are also gene symbols with which this approach falls short.
    A notable example of an extremely polysemic gene symbol is ``CAD'', which represents the gene
    ``carbamoyl-phosphate synthetase 2, aspartate transcarbamylase, and dihydroorotase''.
    The acronym ``CAD'' is ubiquitous in medical literature, where it can be found, among others, standing for:
    \begin{itemize}
        \item Adrenal-demedullated Control
        \item Canine Atopic Dermatitis
        \item Cannabis Dependence
        \item Capillary Area Density
        \item Cervical Artery Dissection
        \item Cinnamyl Alcohol Dehydrogenase
        \item Cold Agglutinin Disease
        \item Collision-Activated Dissociation or Collisionally Activated Dissociation
        \item Computer-Aided Diagnostic
        \item Computer-Assisted Design
        \item Conciliatory Anti-allergic Decoction
        \item Coronary Artery Dilation
        \item Coronary Artery Disease
        \item Coronary Heart Atherosclerotic Disease
    \end{itemize}
    Additionally, ``CAD'' is also the ISO 4217 currency code for the Canadian dollar~\cite{ISO4217}.

    Relying solely on the proximity of gene-related terms, such as ``genes'' and ``genetics'', to extract gene symbols
    like CAD is clearly insufficient\footnote{For example, it systematically mislabels
    articles discussing the genetic aspects of Coronary Artery Disease (CAD) as being about the CAD gene.}.
    However, this approach can significantly reduce the number of matches, making it feasible to supplement the indexing
    process with a semantic analysis of the text using a large language models (LLM) like Meta LLaMA 3 8B~\cite{MetaAI2024}.
    This additional step has proven extremely effective in disambiguating polysemic symbols.

    Disambiguation is achieved by submitting the following prompt to the model with a temperature of 0\footnote{The
    temperature parameter in a Large Language Model (LLM) regulates its output. A lower temperature setting results in
    more deterministic and less creative responses, whereas higher temperatures allow for more varied and imaginative
    outputs.}:
\begin{verbatim}
Answer the following question by true or false. Do not add anything
else.  In the text between single quotes that follows this question,
does the acronym ${symbol} refer to the gene ${geneName}? "${text}"
\end{verbatim}
    Where:
    \begin{itemize}
        \item \$\{symbol\} is replaced by the symbol to extract (e.g., CAD),
        \item \$\{geneName\} is replaced by the name of the gene (e.g., carbamoyl-phosphate synthetase 2, aspartate
        transcarbamylase, and dihydroorotase), and
        \item \$\{text\} is replaced by the title and abstract of the article to index.
    \end{itemize}
    When prompted in this manner, the model returns `true' when the input text contains the gene symbol and `false'
    otherwise\footnote{Note that replacing LLaMA 3 with LLaMA 3.1 resulted in a degradation of result quality,
        characterized by an increased number of cases where the model produced responses not limited to the strict binary
        `true' or `false' expected.}.
    This clear binary response enables straightforward integration of the model call into the data acquisition workflow
    (described in Section~\ref{sec:workflow}).
    Notably, for this specific use case, the results obtained using the 8B version of the model are on par with those
    achieved with the 70B version, while maintaining an acceptable level of latency.
    Moreover, the logic applied:
    \begin{enumerate}
        \item Direct extraction of unambiguous gene symbols (e.g., C11orf30).
        \item Extraction of moderately polysemic symbols (e.g., VHL) when they appear in texts specifically related to genetics.
        \item Application of semantic analysis to disambiguate very polysemic symbols (e.g., CAD).
    \end{enumerate}
    keeps the number of model calls sufficiently low to ensure that these calls do not unreasonably
    increase the overall processing time.
%
%
    \section{PubMed Data Acquisition and Transformation Workflow}
    \label{sec:workflow}
\begin{figure}[ht]
        \centering
        \includegraphics[width=\textwidth]{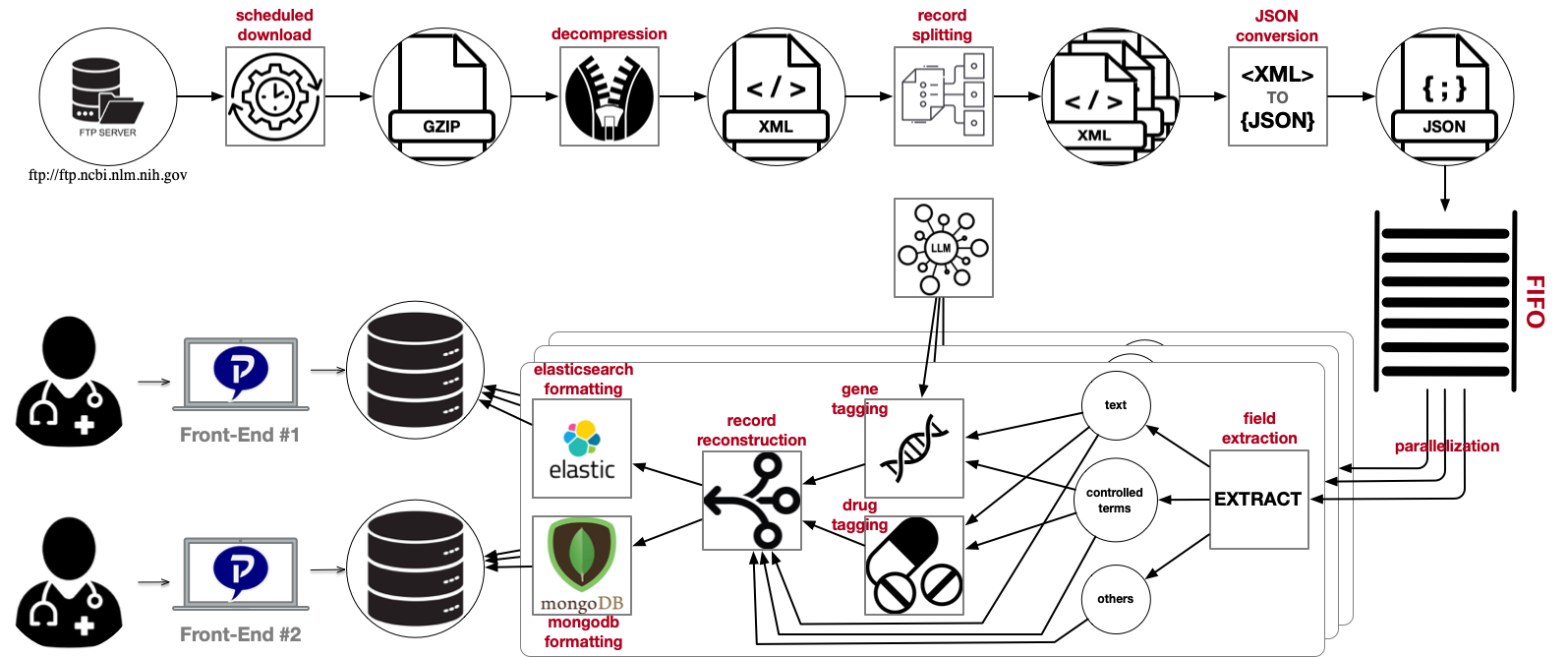}
        \caption{PubMed data acquisition and transformation workflow.}
        \label{fig:workflow}
    \end{figure}

    This section describes the automated workflow that prepares the enriched metadata powering our system.
    The workflow follows a standard Extract, Transform, and Load (ETL) design pattern \cite{Vassiliadis2009}, consisting
    of three main steps:
    \begin{enumerate}
        \item In the extract step, we retrieve raw PubMed records from the National Library of Medicine's (NLM) FTP
        server.
        \item During the transform step, we extract relevant data elements from each record, enrich them as described
        in Section~\ref{sec:indexes}, and format them into the target database schemas.
        \item Finally, in the load step, we insert the enriched records into the target databases.
    \end{enumerate}
    These steps are illustrated in Figure~\ref{fig:workflow} and explained in greater detail below.

    \subsection{Extraction}
    \label{ssec:extract}
    The entire PubMed catalog can be freely obtained from the US National Library of Medicine (NLM) via its FTP server
    at \url{ftp://ftp.ncbi.nlm.nih.gov}.
    The records are available in XML format, packaged in batches of up to 30,000 records, and compressed using the gzip
    standard.
    Daily updates are published at 2pm EST and can be downloaded from the same FTP server.

    The extraction step is relatively straightforward.
    It begins with an initial full extraction, where the entire catalog is downloaded and imported into the data
    pipeline in one go.
    Subsequent updates are then extracted incrementally through scheduled downloads.
    
    \subsection{Transformation}
    \label{ssec:transform}
    After downloading a package, its contents are extracted from the zip file and split into individual XML records.
    These records are then converted into JSON documents and added to a processing queue.

    This allows for easy parallelization of the remaining transformation process.
    Each free processor reads a JSON record from the queue and extracts various types of information it contains,
    including texts (e.g., titles), controlled vocabularies (e.g., MeSH), and other data elements (e.g., dates).

    However, the structure of the JSON documents can vary between records, which complicates the extraction of certain
    data elements.
    For instance, depending on the type of document described, the year of publication or creation can be found in up to
    four different locations.

    Once the relevant data elements are extracted, the texts and controlled terms are passed to a drug indexer module
    and a gene indexer module for further processing, as described in Sections~\ref{ssec:drugs} and~\ref{ssec:genes},
    respectively.
    After these processes complete, the extracted data elements are assembled into their final JSON formats.

    \subsection{Importation}
    \label{ssec:load}

    The transformation workflow produces two types of record formats: one for MongoDB, a document-oriented
    NoSQL database, and another for Elasticsearch, a document-oriented NoSQL search engine.
    This enables seamless loading of the generated records into both systems.

    In both cases, records are uniquely identified by their PubMed identifiers.
    This allows for efficient `upserting' of documents, which means inserting new records and replacing existing ones
    with updated versions.
    This is particularly important because, in addition to publishing descriptions of new documents, the NLM regularly
    releases updates to the descriptions of existing documents.

    Both MongoDB and ElasticSearch can serve as backends for a front-end application called Qibitz, which offers a
    faceted search interface that enables physicians to interactively explore the catalog.
    In the next section, we provide three examples of how this frontend can be leveraged to support various use cases.

%
%
    \section{Identifying Drugs as Potential Therapies for a Given Indication Using Qibitz: Three Examples}
    \label{sec:examples}

    This section reports on three case studies that utilized Qibitz to identify existing drugs with potential for repurposing
    in novel therapeutic areas, where they could be combined with standard treatments for specific indications.
    The process involves two main steps:
    \begin{enumerate}
        \item \textbf{Targeted Literature Search:} Qibitz is used for facetted searching of scientific papers.
        We combine searches for drugs with either mutated genes associated with a specific pathology or MeSH terms
        referencing the same pathology.
        This approach retrieves a comprehensive list of drugs potentially linked to the disease of interest.
        \item \textbf{Prioritization and Refinement:} From the retrieved list, we select drugs identified by Qibitz as
        potential therapies for the given indication based on the following six criteria:
        \begin{itemize}
            \item \textbf{Favorable Safety Profile:} Drugs with a well-established safety profile and minimal risk of
            severe side effects based on their long history of use are preferred.
            \item \textbf{Compatibility and Synergy:} Drugs with no known negative interactions and potential for a
            synergistic effect with standard treatments are prioritized.
            \item \textbf{Demonstrated Long-Term Safety:} Clinical data should support the long-term safety of the
            candidate drug.
            \item \textbf{Broad Pathway Coverage:} Drugs that target multiple pathways relevant to the patient's
            pathology are more promising.
            \item \textbf{Clinically Relevant Dosing:} Preclinical data should suggest a dose range similar to what
            could be used in clinical trials.
            \item \textbf{Existing Clinical Data:} Availability of data from epidemiological studies or previous
            clinical trials involving the drug is advantageous.
        \end{itemize}
    \end{enumerate}
    This approach leverages Qibitz's search capabilities to identify potential drug candidates for repurposing, followed 
    by a data-driven selection process to prioritize the most promising options for further investigation and clinical
    trials.

    \subsection[Advanced Mucosal Melanoma]{Potential Repurposed Drugs Add-on Therapies Identified by Qibitz Based on
    Tumor Genes Alterations Revealed by FoundationOne Next-Generation Sequencing}

    In this first case study, FoundationOne next-generation sequencing (NGS) was applied to the tumor of a 72-year-old
    patient with advanced mucosal melanoma.
    Table~\ref{tab:mutations} lists the cancer-related genes and their mutations detected in the patient's tumor.

    \begin{table}[!htbp]
        \small
        \centering
        \begin{tabular}{|c|c|}
            \hline
            \toprule
            \textbf{Gene} & \textbf{Mutation} \\
            \hline
            \midrule
            \textit{GNAQ} & \textit{Q209P} \\
            \textit{BAP1} & \textit{S680fs*36} \\
            \textit{SF3B1} & \textit{R625C} \\
            \textit{ARID1B} & \textit{G246S} \\
            \textit{HNF1A} & \textit{V590M} \\
            \textit{NOTCH3} & \textit{R113Q} \\
            \textit{ROS1} & \textit{I1231S} \\
            \bottomrule
            \hline
        \end{tabular}
        \caption{Mutated genes found in the patient's tumor.}
        \label{tab:mutations}
    \end{table}

    \paragraph{Mutations and their significance:}

    \begin{itemize}
        \item GNAQ (together with GNA11, not mutated in this tumor) is known as one of the most common driver oncogenes
        in mucosal melanoma.
        Driver oncogenes are genes with mutations that directly promote cancer cell growth and survival.
        \item BAP1 is a tumor suppressor gene.
        Mutations in tumor suppressors can inactivate their function, allowing cancer cells to grow unchecked.
    \end{itemize}

    \subsubsection{Therapeutic implications}

    \subparagraph{GNAQ Q209P Mutation:}
    \begin{itemize}
        \item \textbf{Function:} GNAQ encodes a protein called guanine nucleotide-binding protein G(q) subunit alpha.
        This protein plays a crucial role in cellular signaling pathways.
        \item \textbf{Mutation Significance:} Mutations in GNAQ, particularly Q209P and R183 substitutions, are known
        drivers of mucosal  melanoma.
        These mutations occur in the Ras-like domain and lead to constitutive activation of GNAQ, essentially turning it
        into a cancer-promoting oncogene.
        \item \textbf{Targeted Therapy:} Currently, there are no FDA-approved drugs specifically targeting GNAQ mutations.
        However, preclinical studies suggest that tumors harboring GNAQ mutations might be sensitive to MEK inhibitors.
        \item \textbf{Potential Treatment Options:} Based on this information, FoundationOne proposes FDA-approved MEK
        inhibitors like Binimetinib, Cobimetinib, and Trametinib as potential treatment strategies.
        \item \textbf{Additional Therapeutical Options Suggested by Qibitz:} While PubMed identified 783 articles related
        to "GNAQ", Qibitz narrowed the search by focusing on 50 articles referencing ``GNAQ'' and 44 drugs.
        This analysis revealed additional potential treatments beyond MEK inhibitors, including the AKT/mTOR inhibitor
        Metformin~\cite{AmiroucheneAngelozzi2014} and the antimalarial drugs Chloroquine and
        Hydroxychloroquine~\cite{Truong2020}.
    \end{itemize}

    \subparagraph{BAP1 S680fs*36 Mutation:}

    \begin{itemize}
        \item \textbf{Function:} BAP1 is a tumor suppressor gene with a critical role in regulating various cellular
        processes, including DNA repair and cell cycle control.
        It normally resides within the nucleus of the cell.
        \item \textbf{Mutation Significance:} The S680fs*36 mutation disrupts the BAP1 protein, preventing its proper
        movement into the nucleus.
        This impairs its tumor suppressor function, potentially leading to uncontrolled cell growth and cancer development.
        BAP1 mutations are associated with aggressive disease and poor prognosis in mucosal melanoma.
        \item \textbf{Targeted Therapy:} Currently, there are no FDA-approved drugs specifically targeting BAP1 mutations.
        However, based on BAP1's role in DNA repair, FoundationOne suggests that tumors with BAP1 inactivation might be
        sensitive to drugs that target other DNA repair pathways.
        These include:
        \begin{itemize}
            \item PARP inhibitors: Olaparib.
            \item HDAC inhibitors: Belinostat and Vorinostat.
        \end{itemize}
        \item \textbf{Potential Treatment Options Suggested by Qibitz:} While PubMed identified 1868 articles related to
        ``BAP1'', Qibitz narrowed the search by focusing on 131 articles referencing ``BAP1'' and 66 drugs, identifying
        potentially effective treatment options beyond those suggested by FoundationOne namely the following repurposed
        drugs:
        \begin{itemize}
            \item Valproate \cite{Landreville2012}: An HDAC inhibitor.
            \item Minocycline \cite{Alano2006}: A PARP inhibitor.
        \end{itemize}
        They are known as cheaper and less toxic alternatives to the above-mentioned targeted therapies.
    \end{itemize}

    \subparagraph{SF3B1, ARID1B, HNF1A, NOTCH3, and ROS1:}

    \begin{itemize}
        \item \textbf{Targeted Therapy: }FoundationOne does not propose any potential treatment strategies for the
        mutations of these five genes.
        \item \textbf{Potential Treatment Options Suggested by Qibitz:}
        \begin{itemize}
            \item \textbf{SF3B1:} Tumor-associated mutations in SF3B1 induce a BRCA-like cellular phenotype that confers
            synthetic lethality to DNA-damaging agents and PARP inhibitors, which can be exploited therapeutically.

            While PubMed identified 1262 articles related to ``SF3B1'', Qibitz narrowed the search by focusing on 101
            articles referencing ``SF3B1'' and 56 drugs.
            Qibitz identifies the PARP inhibitor Minocycline~\cite{Furney2013,Lappin2022} as a potential additional
            treatment.
            \item \textbf{ARID1B: }RID1B has been considered a cancer driver gene and is a subunit of the human
            Switch/Sucrose nonfermentable chromatin complex that is mutated in diverse human cancers and regulates
            various biological processes, such as cell development, proliferation, differentiation, DNA replication and
            gene expression.
            Mutations of ARID1B have been identified in lung cancer, papillary thyroid cancer, hepatocellular carcinoma
            and neuroblastoma.

            While PubMed identified 399 articles related to ``ARID1B'', Qibitz narrowed the search by focusing on 8
            articles referencing ``ARID1B'' and 11 drugs.
            Qibitz identifies as a potential additional treatment the following Wnt/b-catenin pathway inhibitors:
            Metformin and Celecoxib~\cite{Vasileiou2015,Ahmed2016}.
            \item \textbf{HNF1A:} Hepatocyte nuclear factor 1A (HNF1A) is the master regulator of liver homeostasis and
            organogenesis and regulates many aspects of hepatocyte functions.
            It acts as a tumor suppressor.
            The V590M mutation is located in the transactivating domain of HNF1A.

            While PubMed identified 2063 articles related to ``HNF1A'', Qibitz narrowed the search by focusing on 556
            articles referencing ``HNF1A'' and 116 drugs.
            Qibitz identifies the AKT/Mtor inhibitor Metformin as potential additional treatment~\cite{Pelletier2010}.
            \item \textbf{NOTCH3:} Overexpression, gene amplification and mis-activation of NOTCH3 are associated with
            different cancers.

            While PubMed identified 2309 articles related to ``NOTCH3'', Qibitz narrowed the search by focusing on 175
            articles referencing ``NOTCH3'' and 108 drugs.
            Qibitz identifies the BRD4 inhibitor Azelastine as potential additional
            treatment~\cite{VillarPrados2019,Wakchaure2019}.
            \item \textbf{ROS1:} While PubMed identified 2564 articles related to ``ROS1'', Qibitz narrowed the search
            by focusing on 481 articles referencing ``ROS1'' and 92 drugs.
            Qibitz identifies the ROS1 inhibitor Crizotinib as potential additional treatment~\cite{Ou2012}.
        \end{itemize}
    \end{itemize}

    \subsubsection{Conclusion}
    FoundationOne identifies several potential treatment strategies in this case.
    For the GNAQ Q209P mutation, they propose FDA-approved MEK inhibitors such as Binimetinib, Cobimetinib, and
    Trametinib.
    Additionally, they suggest BAP1 inactivation might be sensitive to PARP inhibitors (like Olaparib) and HDAC
    inhibitors (including Belinostat and Vorinostat).
    However, FoundationOne does not propose any potential treatment strategies for mutations in SF3B1, ARID1B, HNF1A,
    NOTCH3, and ROS1.

    While FoundationOne offered valuable insights, Qibitz helped identifying additional potential treatment options
    through repurposed drugs for the mucous melanoma patient.
    These include:

    \begin{itemize}
        \item The AKT/mTOR inhibitor Metformin (an antidiabetic drug) for mutations in GNAQ, ARID1B, and HNF1A.
        \item Hydroxychloroquine (used for arthritis) for the GNAQ mutation.
        \item For BAP1 inactivation and SF3B1 mutations, Qibitz suggests the HDAC inhibitor Valproate and the PARP
        inhibitor Minocycline (an antibiotic).
        \item Celecoxib, a COX-2 inhibitor typically used for pain and inflammation, might be beneficial for the
        ARID1B G246S mutation by targeting the Wnt/$\beta$-catenin pathway.
        \item Qibitz also identified Azelastine, an antihistamine H1, as a potential BRD4 inhibitor for the NOTCH3 R113Q
        mutation.
        \item Finally, Crizotinib, typically used for ROS1 mutations, was suggested for the specific ROS1 I1231S mutation.
    \end{itemize}

    After reviewing the proposed drugs through the Medscape Drug Interaction Checker \cite{Medscape2024}, two
    interactions were identified:
    \begin{itemize}
        \item A major interaction exists between Hydroxychloroquine Sulfate and Crizotinib: ``Hydroxychloroquine Sulfate
        and Crizotinib both increase QTc interval and might cause a dangerous abnormal heart rhythm and should not be
        used together''.
        \item A minor interaction was found between Valproic Acid and Celecoxib: ``Valproic acid will increase the level
        or effect of Celecoxib by affecting hepatic enzyme CYP2C9/10 metabolism''.
    \end{itemize}

    \subsection[Parkinson Disease]{Potential Repurposed Drugs Add-on Therapies Identified by Qibitz for Parkinson
    Disease (PD)}

    Parkinson's disease (PD), affecting roughly 1-2\% of individuals over 65, is the second most common neurologic disorder.
    Currently, there is no cure, but treatments can manage symptoms.
    PD presents with both motor and non-motor symptoms.
    Motor symptoms include slowness of movement (bradykinesia), stiffness (rigidity), and tremor at rest.
    Non-motor symptoms can encompass dementia and depression.

    \subsubsection{Therapeutic implications}
    The underlying pathology of PD involves the degeneration of dopamine-producing neurons within the substantia nigra
    pars compacta, a region of the brain.
    Additionally, Lewy bodies, protein clumps primarily composed of $\alpha$-synuclein, are a hallmark feature.
    These alterations disrupt the transmission of dopamine in the nigrostriatal pathway, a circuit critical for movement
    control.

    While PubMed identified 174,150 articles related to ``Parkinson Disease'', Qibitz narrowed the search to 26,495 articles
    referencing 594 drugs.

    Among them, Qibitz identifies:
    \begin{itemize}
        \item \textbf{Melatonin (MLT):} Studies suggest melatonin (MLT) holds promise for PD by alleviating synaptic
        dysfunction and neuroinflammation~\cite{Guo2023,Tchekalarova2023,JimenezDelgado2021}.
        MLT may promote the restoration of dopamine-producing neurons in the substantia nigra pars compacta (SNc) by
        increasing dendritic numbers and restoring synaptic plasticity.
        This neuroprotective effect is believed to be mediated by inhibiting $\alpha$-synuclein aggregation and its
        associated neurotoxicity.
        \item \textbf{Curcumin:} Curcumin has been shown to interact with $\alpha$-synuclein condensates, which are
        linked to PD~\cite{Garodia2023,Wijeweera2023,Xu2022,Patel2022}.
        Recent studies suggest that 4$\alpha$-synuclein undergoes a process called phase separation, which can accelerate
        its aggregation into amyloid fibrils.
        Interestingly, curcumin appears to inhibit this aggregation by interacting with $\alpha$-synuclein condensates
        formed during phase separation.
        \item \textbf{Quercetin:} Quercetin is another promising candidate for PD due to its potential neuroprotective
        effects.
        Studies suggest it may help by efficiently reducing the aggregation of $\alpha$-synuclein protein, a hallmark
        feature of PD associated with neurological decline~\cite{Das2023,Rarinca2023,Wang2023}.
        \item \textbf{Metformin:} Metformin is another potential treatment for PD that may work through several
        mechanisms~\cite{Nowell2023,Agostini2021,Labandeira2022}:
        \begin{enumerate}
            \item \textit{Improving mitochondrial function:} Metformin may activate a pathway involving protein kinase
            B and Nrf2, a cellular regulator that helps protect cells from stress.
            This pathway can improve the function of mitochondria, the energy centers of cells, potentially reducing
            deficits in dopamine-producing neurons.
            \item \textit{Reducing inflammation:} Metformin may dampen inflammation by suppressing the activation of
            microglia, immune cells in the brain, and decreasing the production of inflammatory molecules like
            TNF-$\alpha$ and IL-1$\beta$.
            \item \textit{Protecting neurons:} Metformin might help prevent the loss of dopaminergic neurons, which are
            essential for movement control and are damaged in PD.
            \item \textit{Inhibiting protein aggregation:} Metformin may inhibit the abnormal phosphorylation of
            $\alpha$-synuclein, a protein that forms clumps in the brains of people with PD.
        \end{enumerate}
        \item \textbf{Exenatide:} A study investigated exenatide, a medication typically used for diabetes (glucagon-like
        peptide-1 receptor agonist or GLP-1), as a potential treatment for PD~\cite{Athauda2017,Vijiaratnam2021}.
        In this study, 62 patients with PD were randomly assigned to receive either exenatide or a placebo.

        The researchers evaluated patients' movement abilities while they were not taking any medications
        (off-medication state).
        After 60 weeks, those taking exenatide showed an improvement in their movement scores compared to those on placebo.
        This improvement was statistically significant.
        These findings suggest that exenatide may have lasting positive effects on motor function in PD patients.
        \item \textbf{Vitamin D:} Several studies suggest that maintaining healthy vitamin D levels might be associated
        with a lower risk of developing PD~\cite{Sandeep2023,Lason2023,Suzuki2013}.
        Additionally, research on vitamin D supplementation shows promise.
        Clinical trials indicate that supplementation with vitamin D3 (at a dose of 1200 IU/day for one year) may help
        slow the progression of PD in patients.
        Furthermore, it may also reduce the risk of fractures, a common complication of PD.
        \item \textbf{Modafinil:} Modafinil is a medication that promotes wakefulness by enhancing dopamine signaling
        in the brain.
        Unlike traditional stimulants, modafinil appears to have minimal impact on the extrapyramidal motor system,
        which controls movement.
        This makes it a promising option for managing excessive daytime sleepiness (EDS) in PD patients, a common and
        debilitating symptom.

        Interestingly, a large-scale study called ``Comprehensive Real-World Assessment of Marketed Medications to Guide
        Parkinson's Drug Discovery'' analyzed data on over 117 million people and 2,181 medications.
        The study found that modafinil use was associated with a 54\% reduced risk of developing
        PD~\cite{Generali2014,Cepeda2019}.
        \item \textbf{Minocycline:} Minocycline, a medication typically used for bacterial infections, exhibits various
        properties that might be beneficial for PD. These properties include fighting inflammation, preventing cell
        death, and protecting cells from oxidative stress, all of which play a role in the progression of PD.

        Encouraging results from laboratory and animal studies suggest that minocycline might be a promising candidate
        for neuroprotection in PD. These studies highlight its potential to target multiple pathways involved in the
        neurodegenerative process~\cite{Cankaya2019}.
        \item \textbf{Doxycycline:} Doxycycline has an anti-inflammatory effect in the nervous system, based on
        inhibition of glial activation in the substantia nigra and the striatum and suppression of metalloproteinase
        induction.
        Also, doxycycline reduces the transcription of proinflammatory mediators suppressing the p38 MAPK and NF-$\kappa$B
        signaling pathways.
        Doxycycline inhibits the formation of toxic misfolded forms of $\alpha$-synuclein oligomers through protein aggregation
        and blocks the seeding capacity of preformed aggregates in an in vitro and in vivo
        model~\cite{LopesSantosLobato2023}.
        \item\textbf{N-Acetylcysteine (NAC):} A small study suggests that N-acetylcysteine might directly influence the 
        dopamine system in PD patients, potentially leading to clinical improvements. 
        The study used a technique to measure dopamine activity in the brain called DAT binding.
        Patients with PD who received NAC showed a significant increase (mean increase ranging from 4.4\% to 7.8\%; 
        p $<$ 0.05 for all values) in DAT binding in two brain regions (caudate and putamen) compared to those who did not
        receive NAC. 
        Additionally, patients in the NAC group experienced a significant improvement (mean improvement of 12.9\%, p = 
        0.01) in their UPDRS scores, which is a rating scale for Parkinson's symptoms~\cite{Monti2016}.
        \item \textbf{Zolpidem:} One study analyzed data from two large medical databases, involving nearly 200,000
        Parkinson's patients.
        The researchers employed sophisticated methods to assess the impact of hundreds of drugs on delaying dementia, a
        common complication of PD. This approach identified Zolpidem, a sleep medication, as having a potentially
        beneficial effect on slowing PD progression in both datasets~\cite{Laifenfeld2021}.
        Interestingly, other preliminary studies suggest Zolpidem might also improve motor symptoms in PD patients, even
        with single doses.
        It could even help manage dyskinesias, involuntary movements that can be a side effect of PD
        medications~\cite{Daniele2016}.
        \item \textbf{SGLT2 inhibitors Empagliflozin and Dapagliflozin:} Some oral diabetes medications, including those
        known as SGLT2 inhibitors, GLP-1 receptor agonists, and DPP-4 inhibitors, have shown promise in protecting nerve
        cells in both PD patients and laboratory models~\cite{Lin2021}.
        Additionally, studies suggest that diabetic patients taking these medications may have a lower risk of
        developing PD. Among these, SGLT2 inhibitors, the latest type of medication for diabetes, appear to be
        particularly interesting.
        They might offer neuroprotective effects by improving the function of mitochondria, the cell's power plants,
        and by enhancing the body's antioxidant defenses.
        \item \textbf{Caffeine:} Studies suggest that regular caffeine consumption might be associated with a
        significantly lower risk of developing PD (p $<$ 0.001) and slower disease progression(p = 0.03) in those
        already diagnosed~\cite{Hong2023}.
    \end{itemize}

    \subsubsection{Conclusion}
    PD is a movement disorder caused by a dopamine deficit in the brain.
    Current therapies primarily focus on dopamine modulators or replacements, such as levodopa.
    Although dopamine replacement can help alleviate PD symptoms, therapies targeting the underlying neurodegenerative
    process, (the degeneration of dopamine-producing neurons within the substantia nigra pars compacta upon
    $\alpha$-synuclein aggregation), are limited.
    
    While none of the 14 drugs that Qibitz helped identifying as potential repurposed drugs for PD are part of standard
    of care, some offer promising avenues for further investigation.
    
    Potential benefits of identified drugs:
    \begin{itemize}
        \item Inhibit $\alpha$-synuclein aggregation: Melatonin, curcumin, quercetin, metformin, doxycycline.
        \item Reduce neuroinflammation, oxidative stress, and neurodegeneration: Melatonin, metformin, minocycline,
        doxycycline, SGLT2 inhibitors, GLP-1 receptor agonists, and DPP-4 inhibitors.
        \item Lower risk of developing PD: Modafinil, vitamin D, caffeine, SGLT2 inhibitors, GLP-1 receptor agonists,
        and DPP-4 inhibitors (based on observational studies).
        \item Slow disease progression: Exenatide, modafinil, zolpidem, caffeine, N-acetylcysteine, and Vitamin D
        (supported by clinical trials).
    \end{itemize}

    Synergies between all these non-toxic drugs might be expected without any addition of deleterious side-effects.

    \subsection[Mucoepidermoid Carcinoma]{Potential Repurposed Drugs as Adjunctive Therapies for Mucoepidermoid
    Carcinoma Identified by Qibitz Using Its MeSH Descriptor}

    Mucoepidermoid carcinoma (MEC) accounts for 10-15\% of salivary gland tumors.
    However, its relatively low incidence presents a challenge for conducting large-scale clinical trials and developing
    definitive treatment guidelines. Surgery remains the most common treatment for resectable tumors.
    In addition to surgery, other treatment modalities like chemotherapy, radiation therapy, and immunotherapy are being
    explored for their potential role in managing MEC~\cite{Sama2022}.

    \subsubsection{Therapeutic implications}

    While PubMed identified 4,217 articles related to ``Mucoepidermoid Carcinoma'', Qibitz narrowed the search to 140
    articles referencing 86 drugs.
    Among them, Qibitz identifies:
    \begin{itemize}
        \item \textbf{Tocilizumab:} A humanized anti-human interleukin-6 receptor (IL-6R) antibody, tocilizumab enhances 
        the anti-tumor effect of conventional chemotherapy in preclinical models of mucoepidermoid carcinoma. 
        This suggests that patients might benefit from combination therapy with an IL-6R inhibitor like tocilizumab and 
        a chemotherapeutic agent such as paclitaxel. 
        Emerging evidence suggests a correlation between interleukin-6 signaling and the survival of cancer stem cells, 
        potentially contributing to therapy resistance~\cite{Mochizuki2015}.
        \item \textbf{Antiviral drugs targeting Cytomegalovirus (CMV):} Acyclovir and ganciclovir are being investigated
        due to evidence suggesting that CMV may promote mucoepidermoid carcinoma growth.
        The EGFR $\rightarrow$ ERK pathway offers a promising target, with initial treatment using EGFR inhibitors like
        erlotinib or gefitinib showing tumor regression.
        However, this effect is often limited by the development of drug resistance, leading to tumor regrowth.
        Therefore, long-term tumor control might require combined therapy with both kinase inhibitors and antiviral
        drugs like acyclovir and ganciclovir~\cite{Melnick2013,Melnick2014}.
        \item \textbf{Curcumin:} Curcumin is a NF-$\kappa$ and TGF-$\beta$1 pathways inhibitor.
        It has been reported to exhibit antioxidant, anti-inflammatory, antimicrobial and anticancer activity in vitro
        and in animal models of disease.
        Its anticancer activity is exerted through modulation or inhibition of multiple molecular pathways.
        Notably, curcumin is a potent inhibitor of NF-$\kappa$B, a transcription factor found to play a role in
        tumorigenesis of many human malignancies.

        Studies have shown that TGF-$\beta$1 overexpression at early stages of carcinogenesis provides tumor-suppressive
        effects primarily via growth inhibition, whereas TGF-$\beta$1 overexpression at late stages promotes tumor
        progression, metastasis, and epithelial to mesenchymal transition (EMT), potentially via loss of adhesion
        molecules, angiogenesis, proteinase activation, and immune suppression.
        Several studies have reported the antimetastatic effect of curcumin by its ability to modulate the
        epithelial-to-mesenchymal transition (EMT) process in different cancers and the potential of curcumin in
        mitigating the EMT process induced by TGF-$\beta$1. TGF-$\beta$1 is a key regulator of the invasive and
        metastatic potential of mucoepidermoid carcinoma cells, and patients with this malignancy may benefit from
        targeted therapies blocking effectors of this signaling pathway~\cite{Chung2015,Wang2011,Kandagalla2022}.
        \item \textbf{Valproate:} Valproate is being investigated for its potential as a histone deacetylase inhibitor
        (HDACi) in MEC.
        Interfering with the chromatin organization of tumors using HDAC inhibitors constitutes promising strategies to
        manage MEC.
        Pharmacological inhibition of histone deacetylase interferes with the regulation of multiple oncogenic processes,
        including cancer stem cells (CSC) maintenance.
        Studies have shown that inducing histone acetylation (loosening the chromatin structure) in MEC cells
        significantly impairs the ability of CSCs to survive and function~\cite{CesarSilva2023,Wagner2018}.
        \item \textbf{Rivaroxaban:} Rivaroxaban is a PAR-2 and TGF-$\beta$1 pathways inhibitor.
        Rivaroxaban, a direct inhibitor of coagulation factor Xa, inhibits the PAR-2, (Proteinase-activated receptor 2),
        and TGF-$\beta$1 signaling pathways.
        PAR-2 plays a role in the development and progression of MEC.
        Given the potent pro-invasive and pro-metastatic effects of TGF-$\beta$ and PAR-2, inhibiting PAR2 expression or
        function in the tumoral or stromal compartments of tumors is likely to have a strong impact on metastatic disease.
        This comes on the one hand from inhibition of PAR2-driven (independent of TGF-$\beta$) invasion in response to
        activation by serine proteinases including kallikrein-related peptidases and blood coagulation enzymes of the
        tumor microenvironment and on the other hand from disruption of the PAR2-TGF-$\beta$ crosstalk.
        Moreover, both PAR2 and TGF-$\beta$ can promote angiogenesis through VEGF expression and release and are thus
        essential for tumor survival under hypoxic conditions of the microenvironment.
        PAR2 also maintains a constitutive high level of HIF-1$\alpha$ for angiogenesis promotion and this also explains the
        high propensity for metastatic dissemination of cancer cells in hypoxic
        regions~\cite{Ismerim2021,Witte2016,Zhang2023}.
        \item \textbf{Metformin:} Metformin exerts antiproliferative and anti-metastatic effects via multiple pathways,
        including inactivation of STAT3 and NF-$\kappa$B by inhibition of their nuclear translocation.
        NF-$\kappa$B is a crucial player in several steps of cancer initiation and progression, primarily due to its
        strong antiapoptotic effect in cancer cells.
        In most cell types, NF-$\kappa$B dimers are predominantly inactive in the cytoplasm; however, cancer cells typically
        have high NF-$\kappa$B activity.
        Interestingly, although NF-$\kappa$B is predominantly cytoplasmic in normal salivary gland (NSG), in human MEC
        tumors, all cell lines expressed nuclear NF-$\kappa$B.
        Signal transducers and activators of transcription 3 (STAT3) is one of a family of transcription factors that 
        regulate apoptosis as well as cell proliferation, differentiation, and angiogenesis and a target for inducing 
        apoptosis in various cancers. 
        The inhibition of STAT3 clearly induced apoptosis in human MEC cell lines or tumor
        xenograft~\cite{Wagner2016,Hirsch2013,Saengboonmee2017,Yu2018}.
    \end{itemize}
    
    \subsubsection{Conclusion}

    Salivary gland mucoepidermoid carcinomas (MEC) are the most common salivary malignancy, accounting for 30–35\% of
    all malignant salivary gland tumors.
    Treatment for MEC primarily involves surgery, with radiation therapy often used for high-grade tumors or those with
    positive lymph nodes.
    Unfortunately, high-grade MEC can recur, and these patients often have limited treatment options due to the cancer's
    resistance to chemotherapy.
    Additionally, resistance to radiation can further complicate treatment for advanced-stage MEC.

    A significant hurdle in developing effective therapies for MEC has been the lack of suitable models for research.
    However, recent advancements have led to the characterization of MEC cell lines and patient-derived xenografts.
    These models are finally enabling researchers to investigate the mechanisms of the disease and test potential new
    treatments.

    Due to the lack of financial incentives for specific drug development, MEC relatively low incidence presents
    a challenge for conducting large-scale clinical trials and developing definitive treatment guidelines.
    One promising approach involves the repurposing of existing drugs for use against MEC.
    Qibitz has identified several candidates, including:
    \begin{itemize}
        \item Tocilizumab, an anti-IL-6R antibody, used for arthritis, (interleukin-6 is promoting the survival of MEC
        cancer stem cells, potentially contributing to therapy resistance.
        \item Anti-cytomegalovirus (CMV) drugs like Acyclovir and Gancyclovir, (CMV may promote mucoepidermoid
        carcinoma growth).
        \item Curcumin, a nutraceutical, is a NF-$\kappa$ and TGF-$\beta$1 pathways inhibitor, (NF-$\kappa$B is a
        transcription factor found to play a role in tumorigenesis of many human malignancies and TGF-$\beta$1 is a key
        regulator of the invasive and metastatic potential of MEC).
        \item Valproate, an antiepileptic drug, is a histone deacetylase inhibitor (HDACi), (histone acetylation in MEC
        cells significantly impairs the ability of cancer stem cells to survive and function).
        \item Rivaroxaban, an anticoagulant, is a PAR-2, (Proteinase-activated receptor 2), and TGF-beta1 pathways
        inhibitor, (PAR-2 plays a role in the development and progression of MEC).
        \item Metformin, an antidiabetic drug, exerts antiproliferative and anti-metastatic effects by targeting STAT3
        \item and NF-$\kappa$B, (STAT3 inhibition clearly induced apoptosis in human MEC cell lines or tumor xenograft).
    \end{itemize}

    Synergy is expected between these repurposed drugs targeting complementary drivers pathways of activation.

%
%
    \section*{Conclusion}
    \addcontentsline{toc}{section}{Conclusion}

    In this paper, we utilized PubMed, a comprehensive catalog of medical literature, and restructured its information
    content, primarily the MeSH descriptors, supplemented by free text data extraction to enrich the catalog with two
    new specialized indexes: one for drugs and another for genes.
    We then developed a faceted search interface to leverage these new indexes and interact with the catalog's content
    as a whole, revealing emerging patterns from the entire corpus of literature metadata.

    These metadata patterns enabled us to acquire new knowledge about the potential of repurposing existing drugs to
    contribute to the treatment of specific diseases including cancers presenting particular gene mutations.
    We illustrated this method by applying it to three specific cases:
    \begin{enumerate}
        \item Identifying potential repurposed drugs as add-on therapies based on tumor gene alterations revealed by
        FoundationOne Next-Generation Sequencing,
        \item Identifying potential repurposed drugs as add-on therapies for Parkinson's disease, and
        \item Identifying potential repurposed drugs as adjunctive therapies for Mucoepidermoid Carcinoma.
    \end{enumerate}
    Each case study led to the suggestion of candidate treatments that can be tested in future clinical trials.

    Notably, the methodology enabled by our tool has already led to the development of ``Polypill'' a multi-ingredient
    pharmaceutical composition patented in 2019 for use in cancer therapy~\cite{Zeicher2019}.

    The two specialized indexes, one for drugs and one for genes (sourced from FoundationOne's list of cancer-related genes),
    enable drug searches by gene or MeSH descriptor.
    This capability allows us to identify potential repurposed drugs for various health conditions, both pathological
    and non-pathological (such as longevity, senolytic activity, cognitive status), provided a corresponding MeSH
    descriptor exists in PubMed.
    Moreover, our approach can identify genes influenced by specific drugs or associated with particular health conditions.

    In an era where Big Data and advanced AI are employed to discover costly new treatments that may take over a decade
    to reach the market, this paper presents a complementary methodology and supporting tool that empowers physicians
    to utilize existing data from published research literature.
    This approach enables the identification of combinations of potentially repurposable drugs with well-studied side
    effects that can be tested in shorter and more cost-effective clinical trials.


    \newpage
    \addcontentsline{toc}{section}{References}
    \bibliography{qibitz}

\begin{thebibliography}{10}

\bibitem{ISO4217}
Currency codes.
\newblock Standard ISO 4217:2015, International Organization for
  Standardization, Geneva, Switzerland, 2015.

\bibitem{Agostini2021}
F.~Agostini, A.~Masato, L.~Bubacco, and M.~Bisaglia.
\newblock Metformin repurposing for {P}arkinson disease therapy: Opportunities
  and challenges.
\newblock {\em International Journal of Molecular Sciences}, 23(1):398, January
  2022.

\bibitem{Ahmed2016}
K.~Ahmed, H.~Shaw, A.~Koval, and V.~Katanaev.
\newblock A second {WNT} for old drugs: Drug repositioning against
  {WNT}-dependent cancers.
\newblock {\em Cancers (Basel)}, 8(7):66, July 2016.

\bibitem{Alano2006}
C.~Alano, T.~Kauppinen, A.~Viader~Valls, and R.~Swanson.
\newblock Minocycline inhibits poly({ADP}-ribose) polymerase-1 at nanomolar
  concentrations.
\newblock In {\em Proceedings of the National Academy of Sciences of the United
  States of America}, volume 103, pages 9685--9690, June 2006.

\bibitem{AmiroucheneAngelozzi2014}
N.~Amirouchene-Angelozzi, F.~Nemati, D.~Gentien, A.~Nicolas, A.~Dumonte,
  G.~Carita, J.~Camonis, L.~Desjardins, N.~Cassoux, S.~Piperno-Neumann,
  P.~Mariani, X.~Sastre, D.~Decaudin, and S.~Roman-Roman.
\newblock Establishment of novel cell lines recapitulating the genetic
  landscape of uveal melanoma and preclinical validation of m{TOR} as a
  therapeutic target.
\newblock {\em Molecular Oncology}, 8(8):1508--1520, December 2014.

\bibitem{Ashburn2004}
T.~Ashburn and K.~Thor.
\newblock Drug repositioning: identifying and developing new uses for existing
  drugs.
\newblock {\em Nature Reviews Drug Discovery}, 3(8):673--683, 2004.

\bibitem{Athauda2017}
D.~Athauda, K.~Maclagan, S.~Skene, M.~Bajwa-Joseph, D.~Letchford, K.~Chowdhury,
  S.~Hibbert, N.~Budnik, L.~Zampedri, J.~Dickson, Y.~Li, I.~Aviles-Olmos,
  T.~Warner, P.~Limousin, A.~Lees, N.~Greig, S.~Tebbs, and T.~Foltynie.
\newblock Exenatide once weekly versus placebo in {P}arkinson's disease: a
  randomised, double-blind, placebo-controlled trial.
\newblock {\em Lancet}, 390(10103):1664--1675, October 2017.

\bibitem{Cankaya2019}
S.~Cankaya, B.~Cankaya, U.~Kilic, E.~Kilic, and B.~Yulug.
\newblock The therapeutic role of minocycline in {P}arkinson's disease.
\newblock {\em Drugs in Context}, 8, March 2019.

\bibitem{Cepeda2019}
M.~Cepeda, D.~Kern, G.~Seabrook, and S.~Lovestone.
\newblock Comprehensive real-world assessment of marketed medications to guide
  {P}arkinson's drug discovery.
\newblock {\em Clinical Drug Investigation}, 39(11):1067--1075, November 2019.

\bibitem{CesarSilva2023}
L.~C{\'e}sar~Silva, G.~Bonif{\'a}cio~Borgato, V.~Petersen~Wagner,
  M.~Domingues~Martins, M.~Ajudarte~Lopes, A.~Santos-Silva, G.~De~Castro,
  L.~Kowalski, C.~Squarize, P.~Vargas, and R.~Moraes~Castilho.
\newblock Repurposing {NF}$\kappa${B} and {HDAC} inhibitors to individually
  target cancer stem cells and non-cancer stem cells from mucoepidermoid
  carcinomas.
\newblock {\em American Journal of Cancer Research}, 13(4):1547--1559, April
  2023.

\bibitem{Chung2015}
S.~Chung and J.~Vadgama.
\newblock Curcumin and epigallocatechin gallate inhibit the cancer stem cell
  phenotype via down-regulation of {STAT3-NF}$\kappa${B} signaling.
\newblock {\em Anticancer Research}, 35(1):39--46, January 2015.

\bibitem{Daniele2016}
A.~Daniele, F.~Panza, A.~Greco, G.~Logroscino, and D.~Seripa.
\newblock Can a positive allosteric modulation of {GABA}ergic receptors improve
  motor symptoms in patients with {P}arkinson's disease? the potential role of
  zolpidem in the treatment of {P}arkinson's disease.
\newblock {\em Parkinson's Disease}, May 2016.

\bibitem{FAMHP2023}
{Federal Agency for Medicines and Health Products}.
\newblock Database of authorized medicines for human use.
\newblock \url{https://medicinesdatabase.be/human-use}, 2023.
\newblock Accessed July 1, 2023.

\bibitem{FoundationOne2023}
{Foundation Medicine}.
\newblock {FoundationOne{\textregistered}CDx}.
\newblock \url{https://www.foundationmedicine.com/test/foundationone-cdx}.
\newblock Accessed July 4, 2023.

\bibitem{Frampton2013}
G.~M. Frampton, A.~Fichtenholtz, G.~A. Otto, K.~Wang, S.~R. Downing, J.~He,
  M.~Schnall-Levin, J.~White, E.~M. Sanford, P.~An, J.~Sun, F.~Juhn,
  K.~Brennan, K.~Iwanik, A.~Maillet, J.~Buell, E.~White, M.~Zhao,
  S.~Balasubramanian, S.~Terzic, T.~Richards, V.~Banning, L.~Garcia,
  K.~Mahoney, Z.~Zwirko, A.~Donahue, H.~Beltran, J.~M. Mosquera, M.~A. Rubin,
  S.~Dogan, C.~V. Hedvat, M.~F. Berger, L.~Pusztai, M.~Lechner, C.~Boshoff,
  M.~Jarosz, C.~Vietz, A.~Parker, V.~A. Miller, J.~S. Ross, J.~Curran, M.~T.
  Cronin, P.~J. Stephens, D.~Lipson, and R.~Yelensky.
\newblock Development and validation of a clinical cancer genomic profiling
  test based on massively parallel {DNA} sequencing.
\newblock {\em Nature Biotechnology}, 31(11):1023--1031, 2013.

\bibitem{Furney2013}
S.~Furney, M.~Pedersen, D.~Gentien, A.~Dumont, A.~Rapinat, L.~Desjardins,
  S.~Turajlic, S.~Piperno-Neumann, P.~{de la Grange}, S.~Roman-Roman, M.-H.
  Stern, and R.~Marais.
\newblock {SF3B1} mutations are associated with alternative splicing in uveal
  melanoma.
\newblock {\em Cancer Discovery}, 3(10):1122--1129, October 2013.

\bibitem{Garodia2023}
P.~Garodia, M.~Hegde, A.~Kunnumakkara, and B.~Aggarwal.
\newblock Curcumin, inflammation, and neurological disorders: How are they
  linked?
\newblock {\em Integrative Medicine Research}, 12(3), September 2023.

\bibitem{Generali2014}
J.~Generali and D.~Cada.
\newblock Modafinil: Parkinson disease--related somnolence.
\newblock {\em Hospital Pharmacy}, 49(7):612--614, July 2014.

\bibitem{Guo2023}
Y.-L. Guo, X.-J. Wei, T.~Zhang, and T.~Sun.
\newblock Molecular mechanisms of melatonin-induced alleviation of synaptic
  dysfunction and neuroinflammation in {P}arkinson's disease: A review.
\newblock {\em European Review for Medical and Pharmacological Sciences},
  27(11):5070--5082, June 2023.

\bibitem{Heguy1995}
A.~Heguy, A.~Stewart, J.~Haley, D.~Smith, and J.~Foulkes.
\newblock Gene expression as a target for new drug discovery.
\newblock {\em Gene Expression}, 4(6):337--344, 1995.

\bibitem{Hirsch2013}
H.~Hirsch, D.~Iliopoulos, and K.~Struhl.
\newblock Metformin inhibits the inflammatory response associated with cellular
  transformation and cancer stem cell growth.
\newblock In {\em Proceedings of the National Academy of Sciences of the United
  States of America}, volume 110, pages 972--977, January 2013.

\bibitem{Hong2023}
C.~Hong, L.~Chan, and C.-H. Bai.
\newblock The effect of caffeine on the risk and progression of {P}arkinson's
  disease: A meta-analysis.
\newblock {\em Nutrients}, 15(3):699, January 2023.

\bibitem{Ismerim2021}
A.~Ismerim, I.~de~Oliveira~Ara{\'u}jo, F.~de~Aquino~Xavier, C.~Rocha,
  C.~Macedo, M.~Cangussu, V.~Freitas, R.~Della~Coletta, P.~Cury, and J.~Santos.
\newblock Mast cells and proteins related to myofibroblast differentiation
  ({PAR-2}, {IL-6}, and {TGF}$\beta$1) in salivary cancers: A preliminary
  study.
\newblock {\em Applied Immunohistochemistry \& Molecular Morphology (AIMM)},
  29(7):e57--e67, August 2021.

\bibitem{JimenezDelgado2021}
A.~Jim{\'e}nez-Delgado, G.~Ortiz, D.~Delgado-Lara, H.~Gonz{\'a}lez-Usigli,
  L.~Gonz{\'a}lez-Ortiz, M.~Cid-Hern{\'a}ndez, J.~Cruz-Serrano, and
  F.~Pacheco-Mois{\'e}s.
\newblock Effect of melatonin administration on mitochondrial activity and
  oxidative stress markers in patients with {P}arkinson's disease.
\newblock {\em Oxidative Medicine and Cellular Longevity}, October 2021.

\bibitem{Kandagalla2022}
S.~Kandagalla, B.~Sharath, A.~Sherapura, M.~Grishina, V.~Potemkin, J.~Lee,
  G.~Ramaswamy, B.~Prabhakar, and M.~Hanumanthappa.
\newblock A systems biology investigation of curcumin potency against
  {TGF}-$\beta$-induced {EMT} signaling in lung cancer.
\newblock {\em 3 Biotech}, 12(11):306, November 2022.

\bibitem{Labandeira2022}
C.~Labandeira, A.~Fraga-Bau, D.~Arias~Ron, E.~Alvarez-Rodriguez,
  P.~Vicente-Alba, J.~Lago-Garma, and A.~Rodriguez-Perez.
\newblock Parkinson's disease and diabetes mellitus: Common mechanisms and
  treatment repurposing.
\newblock {\em Neural Regeneration Research}, 17(8):1652--1658, August 2022.

\bibitem{Laifenfeld2021}
D.~Laifenfeld, C.~Yanover, M.~Ozery-Flato, O.~Shaham, M.~Rosen-Zvi, N.~Lev,
  Y.~Goldschmidt, and I.~Grossman.
\newblock Emulated clinical trials from longitudinal real-world data
  efficiently identify candidates for neurological disease modification:
  Examples from {P}arkinson's disease.
\newblock {\em Frontiers in Pharmacology}, 12, April 2021.

\bibitem{Landreville2012}
S.~Landreville, O.~Agapova, K.~Matatall, Z.~Kneass, M.~Onken, R.~Lee,
  A.~Bowcock, and J.~Harbour.
\newblock Histone deacetylase inhibitors induce growth arrest and
  differentiation in uveal melanoma.
\newblock {\em Clinical Cancer Research}, 18(2):408--416, October 2012.

\bibitem{Lappin2022}
K.~Lappin, E.~Barros, S.~Jhujh, G.~Irwin, H.~McMillan, F.~Liberante,
  C.~Latimer, M.~{La Bonte}, K.~Mills, D.~Harkin, G.~Stewart, and K.~I. Savage.
\newblock Cancer-associated {SF3B1} mutations confer a {BRCA}-like cellular
  phenotype and synthetic lethality to {PARP} inhibitors.
\newblock {\em Cancer Research}, 82(5):819--830, March 2022.

\bibitem{Lason2023}
W.~Laso{\'n}, D.~Jantas, M.~Le{\'s}kiewicz, M.~Regulska, and A.~Basta-Kaim.
\newblock The vitamin {D} receptor as a potential target for the treatment of
  age-related neurodegenerative diseases such as {A}lzheimer's and
  {P}arkinson's diseases: A narrative review.
\newblock {\em Cells}, 12(4):660, February 2023.

\bibitem{Lin2021}
K.-J. Lin, T.-J. Wang, S.-D. Chen, K.-L. Lin, C.-W. Liou, M.-Y. Lan, Y.-C.
  Chuang, J.-H. Chuang, P.-W. Wang, J.-J. Lee, F.-S. Wang, H.-Y. Lin, and T.-K.
  Lin.
\newblock Two birds one stone: The neuroprotective effect of antidiabetic
  agents on {P}arkinson disease --- focus on sodium-glucose cotransporter 2
  ({SGLT2}) inhibitors.
\newblock {\em Antioxidants}, 10(12):1935, December 2021.

\bibitem{LopesSantosLobato2023}
B.~Lopes Santos-Lobato, M.~Capellari Macruz~Brito, {\^A}.~Vieira~Pimentel,
  R.~Torres Oliveira~Cavalcanti, E.~Del-Bel, and V.~Tumas.
\newblock Doxycycline to treat levodopa-induced dyskinesias in {P}arkinson's
  disease: A preliminary study.
\newblock {\em Arquivos de Neuro-Psiquiatria}, 81(5):460--468, May 2023.

\bibitem{Medscape2024}
Medscape.
\newblock Drug interaction checker.
\newblock \url{https://reference.medscape.com/drug-interactionchecker}, 2024.
\newblock Accessed April 14, 2024.

\bibitem{Melnick2014}
M.~Melnick, K.~Deluca, and T.~Jaskoll.
\newblock {CMV}-induced pathology: Pathway and gene-gene interaction analysis.
\newblock {\em Experimental and Molecular Pathology}, 97(1):154--165, August
  2014.

\bibitem{Melnick2013}
M.~Melnick, P.~Sedghizadeh, K.~Deluca, and T.~Jaskoll.
\newblock Cytomegalovirus-induced salivary gland pathology: resistance to
  kinase inhibitors of the upregulated host cell {EGFR/ERK} pathway is
  associated with {CMV}-dependent stromal overexpression of {IL-6} and
  fibronectin.
\newblock {\em Herpesviridae}, 4(1), January 2013.

\bibitem{MetaAI2024}
{Meta AI}.
\newblock Introducing {M}eta {L}lama 3: The most capable openly available {LLM}
  to date.
\newblock \url{https://ai.meta.com/blog/meta-llama-3/}, April 2024.
\newblock Accessed July 14, 2024.

\bibitem{Mochizuki2015}
D.~Mochizuki, A.~Adams, K.~Warner, Z.~Zhang, A.~Pearson, K.~Misawa, S.~McLean,
  G.~Wolf, and J.~N{\"o}r.
\newblock Anti-tumor effect of inhibition of {IL-6} signaling in mucoepidermoid
  carcinoma.
\newblock {\em Oncotarget}, 6(26):22822---22835, September 2015.

\bibitem{Monti2016}
D.~Monti, G.~Zabrecky, D.~Kremens, T.-W. Liang, N.~Wintering, J.~Cai, X.~Wei,
  A.~Bazzan, L.~Zhong, B.~Bowen, C.~Intenzo, L.~Iacovitti, and A.~B. Newberg.
\newblock N-acetyl cysteine may support dopamine neurons in {P}arkinson's
  disease: Preliminary clinical and cell line data.
\newblock {\em PLoS One}, 11(6), June 2016.

\bibitem{PubMed2023}
{National Library of Medicine}.
\newblock About {PubMed}.
\newblock \url{https://pubmed.ncbi.nlm.nih.gov/about/}.
\newblock Accessed April 12, 2023.

\bibitem{HGNC}
{National Library of Medicine}.
\newblock {HGNC} ({HUGO} {G}ene {N}omenclature {C}ommittee) - {S}ynopsis.
\newblock
  \url{https://www.nlm.nih.gov/research/umls/sourcereleasedocs/current/HGNC/index.html}.
\newblock Accessed July 4, 2023.

\bibitem{MeSH2023}
{National Library of Medicine}.
\newblock {Medical Subject Headings (MeSH)}.
\newblock \url{https://www.nlm.nih.gov/mesh/meshhome.html}.
\newblock Accessed April 12, 2023.

\bibitem{PubMedFields2023}
{National Library of Medicine}.
\newblock {MEDLINE/PubMed Data Element (Field) Descriptions}.
\newblock \url{https://www.nlm.nih.gov/bsd/mms/medlineelements.html}.
\newblock Accessed April 12, 2023.

\bibitem{Nowell2023}
J.~Nowell, E.~Blunt, D.~Gupta, and P.~Edison.
\newblock Antidiabetic agents as a novel treatment for {A}lzheimer's and
  {P}arkinson's disease.
\newblock {\em Ageing Research Reviews}, August 2023.

\bibitem{PubMedGuide2023}
{Ohio State University}.
\newblock {What is PubMed?}
\newblock \url{https://hslguides.osu.edu/pubmed/what-is-pubmed}.
\newblock Accessed April 12, 2023.

\bibitem{Ou2012}
S.-H. Ou, C.~Huang~Bartlett, M.~Mino-Kenudson, J.~Cui, and A.~Iafrate.
\newblock Crizotinib for the treatment of {ALK}-rearranged non-small cell lung
  cancer: A success story to usher in the second decade of molecular targeted
  therapy in oncology.
\newblock {\em The Oncologist}, 17(11):1351--1375, September 2012.

\bibitem{Patel2022}
A.~Patel, C.~Olang, G.~Lewis, K.~Mandalaneni, N.~Anand, and V.~Gorantla.
\newblock An overview of {P}arkinson's disease: Curcumin as a possible
  alternative treatment.
\newblock {\em Cureus}, 14(5), May 2022.

\bibitem{Pelletier2010}
L.~Pelletier, S.~Rebouissou, A.~Paris, E.~Rathahao-Paris, E.~Perdu,
  P.~Bioulac-Sage, S.~Imbeaud, and J.~Zucman-Rossi.
\newblock Loss of hepatocyte nuclear factor 1alpha function in human
  hepatocellular adenomas leads to aberrant activation of signaling pathways
  involved in tumorigenesis.
\newblock {\em Hepatology (Baltimore, Md.)}, 51(2):557--566, February 2010.

\bibitem{Rarinca2023}
V.~Rarinca, M.~Nicoara, D.~Ureche, and A.~Ciobica.
\newblock Exploitation of quercetin's antioxidative properties in potential
  alternative therapeutic options for neurodegenerative diseases.
\newblock {\em Antioxidants}, 12(7), July 2023.

\bibitem{Das2023}
S.~Sachi~Das, N.~Kumar~Jha, S.~Kumar~Jha, P.~Ranjan Prasad~Verma, G.~Ashraf,
  and S.~Singh.
\newblock Neuroprotective role of quercetin against alpha-synuclein-associated
  hallmarks in {P}arkinson's disease.
\newblock {\em Current Neuropharmacology}, 21(7):1464--1466, 2023.

\bibitem{Saengboonmee2017}
C.~Saengboonmee, W.~Seubwai, U.~Cha'on, K.~Sawanyawisuth, S.~Wongkham, and
  C.~Wongkham.
\newblock Metformin exerts antiproliferative and anti-metastatic effects
  against cholangiocarcinoma cells by targeting {STAT3} and {NF}-$\kappa${B}.
\newblock {\em Anticancer Research}, 37(1):115--123, January 2017.

\bibitem{Sama2022}
S.~Sama, T.~Komiya, and A.~Guddati.
\newblock Advances in the treatment of mucoepidermoid carcinoma.
\newblock {\em World Journal of Oncology}, 13(1):1--7, February 2022.

\bibitem{Sandeep2023}
Sandeep, M.~Sahu, L.~Rani, A.~Kharat, and A.~Mondal.
\newblock Could vitamins have a positive impact on the treatment of
  {P}arkinson's disease?
\newblock {\em Brain Sciences}, 13(2):272, February 2023.

\bibitem{Suzuki2013}
M.~Suzuki, M.~Yoshioka, M.~Hashimoto, M.~Murakami, M.~Noya, D.~Takahashi, and
  M.~Urashima.
\newblock Randomized, double-blind, placebo-controlled trial of vitamin {D}
  supplementation in {P}arkinson disease.
\newblock {\em The American Journal of Clinical Nutrition}, 97(5):1004--1013,
  May 2013.

\bibitem{Tchekalarova2023}
J.~Tchekalarova and R.~Tzoneva.
\newblock Oxidative stress and aging as risk factors for {A}lzheimer's disease
  and {P}arkinson's disease: The role of the antioxidant melatonin.
\newblock {\em International Journal of Molecular Sciences}, 24(3):3022,
  February 2023.

\bibitem{Truong2020}
A.~Truong, J.~Yoo, M.~Scherzer, J.~Sanchez, K.~Dale, C.~Kinsey, J.~Richards,
  D.~Shin, P.~Ghazi, M.~Onken, K.~Blumer, S.~Odelberg, and M.~McMahon.
\newblock Chloroquine sensitizes {GNAQ}/11-mutated melanoma to {MEK}1/2
  inhibition.
\newblock {\em Clinical Cancer Research}, 26(23):6374--6386, December 2020.

\bibitem{Tunkelang2009}
D.~Tunkelang.
\newblock {\em Faceted Search}.
\newblock Synthesis Lectures on Information Concepts, Retrieval, and Services.
  Morgan and Claypool Publishers, 2009.

\bibitem{Vasileiou2015}
G.~Vasileiou, A.~Ekici, S.~Uebe, C.~Zweier, J.~Hoyer, H.~Engels, J.~Behrens,
  A.~Reis, and M.~Hadjihannas.
\newblock Chromatin-remodeling-factor {ARID1B} represses
  {W}nt/$\beta$-{C}atenin signaling.
\newblock {\em American Journal of Human Genetics}, 97(3):445--456, September
  2015.

\bibitem{Vassiliadis2009}
P.~Vassiliadis, A.~Simitsis, and E.~Baikousi.
\newblock A taxonomy of {ETL} activities.
\newblock In {\em DOLAP '09: Proceedings of the ACM twelfth international
  workshop on Data warehousing and OLAP}, pages 25--32. Association for
  Computing Machinery, November 2009.

\bibitem{Vijiaratnam2021}
N.~Vijiaratnam, C.~Girges, G.~Auld, M.~Chau, K.~Maclagan, A.~King, S.~Skene,
  K.~Chowdhury, S.~Hibbert, H.~Morris, P.~Limousin, D.~Athauda, C.~Carroll,
  M.~Hu, M.~Silverdale, G.~Duncan, R.~Chaudhuri, C.~Lo, S.~Del~Din, A.~Yarnall,
  L.~Rochester, R.~Gibson, J.~Dickson, R.~Hunter, V.~Libri, and T.~Foltynie.
\newblock Exenatide once weekly over 2 years as a potential disease-modifying
  treatment for {P}arkinson's disease: {P}rotocol for a multicentre,
  randomised, double blind, parallel group, placebo controlled, phase 3 trial:
  {T}he '{Exenatide-PD3}' study.
\newblock {\em BMJ Open}, 11(5), May 2021.

\bibitem{VillarPrados2019}
A.~Villar-Prados, S.~Wu, K.~Court, S.~Ma, C.~LaFargue, M.~Chowdhury,
  M.~Engelhardt, C.~Ivan, P.~Ram, Y.~Wang, K.~Baggerly, C.~Rodriguez-Aguayo,
  G.~Lopez-Berestein, S.~Ming-Yang, D.~Maloney, M.~Yoshioka, J.~Strovel,
  J.~Roszik, and A.~Sood.
\newblock Predicting novel therapies and targets: Regulation of notch3 by the
  bromodomain protein {BRD4}.
\newblock {\em Molecular Cancer Therapeutics}, 18(2):421--436, February 2019.

\bibitem{Wagner2018}
V.~Wagner, M.~Martins, M.~Martins, L.~Almeida, K.~Warner, J.~N{\"o}r,
  C.~Squarize, and R.~Castilho.
\newblock Targeting histone deacetylase and {NF}$\kappa${B} signaling as a
  novel therapy for mucoepidermoid carcinomas.
\newblock {\em Scientific Reports}, 8(1):2065, February 2018.

\bibitem{Wagner2016}
V.~Wagner, M.~Martins, M.~Martins, K.~Warner, L.~Webber, C.~Squarize,
  J.~N{\"o}r, and R.~Castilho.
\newblock Overcoming adaptive resistance in mucoepidermoid carcinoma through
  inhibition of the {IKK}-$\beta$/i$\kappa$ba$\kappa$/{NF}$\kappa${B} axis.
\newblock {\em Oncotarget}, 7(45):73032--73044, November 2016.

\bibitem{Wakchaure2019}
P.~Wakchaure, R.~Velayutham, and K.~Roy.
\newblock Structure investigation, enrichment analysis and structure-based
  repurposing of {FDA}-approved drugs as inhibitors of {BET-BRD4}.
\newblock {\em Journal of Biomolecular Structure \& Dynamics},
  37(12):3048--3057, November 2019.

\bibitem{Wang2011}
J.~Wang, J.~Chen, K.~Zhang, Y.~Zhao, J.~N{\"o}r, and J.~Wu.
\newblock {TGF}-$\beta$1 regulates the invasive and metastatic potential of
  mucoepidermoid carcinoma cells.
\newblock {\em Journal of Oral Pathology \& Medicine}, 40(10):762--768,
  November 2011.

\bibitem{Wang2023}
P.~Wang, Q.~Chen, Z.~Tang, L.~Wang, B.~Gong, M.~Li, S.~Li, and M.~Yang.
\newblock Uncovering ferroptosis in {P}arkinson's disease via bioinformatics
  and machine learning, and reversed deducing potential therapeutic natural
  products.
\newblock {\em Frontiers in Genetics}, July 2023.

\bibitem{Wijeweera2023}
G.~Wijeweera, N.~Wijekoon, L.~Gonawala, Y.~Imran, C.~Mohan, and K.~De~Silva.
\newblock Therapeutic implications of some natural products for neuroimmune
  diseases: A narrative of clinical studies review.
\newblock {\em Evidence-Based Complementary and Alternative Medicine}, 2023,
  April 2023.

\bibitem{Witte2016}
D.~Witte, F.~Zeeh, T.~G{\"a}deken, F.~Gieseler, B.~Rauch, U.~Settmacher,
  R.~Kaufmann, H.~Lehnert, and H.~Ungefroren.
\newblock Proteinase-activated receptor 2 is a novel regulator of {TGF}-$\beta$
  signaling in pancreatic cancer.
\newblock {\em Journal of Clinical Medicine}, 5(12):111, December 2016.

\bibitem{Xu2022}
B.~Xu, J.~Chen, and Y.~Liu.
\newblock Curcumin interacts with $\alpha$-synuclein condensates to inhibit
  amyloid aggregation under phase separation.
\newblock {\em ACS Omega}, 7(34):29526--30656, August 2022.

\bibitem{Yu2018}
H.-J. Yu, C.-H. Ahn, I.-H. Yang, D.-H. Won, B.~Jin, N.-P. Cho, S.~Hong, J.-A.
  Shin, and S.-D. Cho.
\newblock Apoptosis induced by methanol extract of potentilla discolor in human
  mucoepidermoid carcinoma cells through {STAT3/PUMA} signaling axis.
\newblock {\em Molecular Medicine Reports}, 17(4):5258--5264, April 2018.

\bibitem{Zeicher2019}
M.~Zeicher.
\newblock Multi-ingredient pharmaceutical composition for use in cancer
  therapy.
\newblock U.S. Patent US-10413558-B2, September 2019.

\bibitem{Zhang2023}
Q.~Zhang, Z.~Zhang, W.~Chen, H.~Zheng, D.~Si, and W.~Zhang.
\newblock Rivaroxaban, a direct inhibitor of coagulation factor {Xa},
  attenuates adverse cardiac remodeling in rats by regulating the {PAR-2} and
  {TGF}-$\beta$1 signaling pathways.
\newblock {\em PeerJ}, September 2023.

\end{thebibliography}
    \bibliographystyle{abbrv}


\end{document}